\documentclass[runningheads]{llncs}
\usepackage[T1]{fontenc}
\usepackage{graphicx}
\usepackage{color}

\newtheorem{hypothesis}{Hypothesis}
\usepackage{enumitem}
\usepackage{amsmath}    
\usepackage{amssymb}    
\usepackage{amsfonts}   

\begin{document}
\title{Strategic Behavior in Crowdfunding: Insights from a Large-Scale Online Experiment.}
%
%
\author{Din Amir\inst{1}
\and
Bar Hoter\inst{1}
\and
Moran Koren\inst{1}\orcidID{0000-0003-0012-0208}}
\authorrunning{F. Author et al.}
%
\institute{Ben-Gurion University of the Negev, Department of Industrial Engineering and Management} 
%
\maketitle              
\begin{abstract}
This study examines strategic behavior in crowdfunding using a large-scale online experiment. Building  on  the model of \cite{arieli2023information}, we test predictions about risk aversion (i.e., opting out despite seeing a positive private signal) and mutual insurance (i.e., opting in despite seeing a negative private signal) in a static, single-shot crowdfunding game, focusing on informational incentives rather than dynamic effects. Our results validate key theoretical predictions:   crowdfunding mechanisms induce distinct strategic behaviors compared to voting, where participants are more likely to follow private signals (odds ratio: 0.139, $p < 0.001$). Additionally, the study demonstrates that higher signal accuracy (85\% vs. 55\%) decreases risk aversion (odds ratio: 0.414, $p = 0.024$) but increases reliance on mutual insurance (odds ratio: 2.532, $p = 0.026$). However, contrary to theory, increasing the required participation  threshold  (50\% to 80\%) amplifies risk aversion (odds ratio: 3.251, $p = 0.005$), which, pending further investigation, may indicate cognitive constraints.  
 Furthermore, we show that while mutual insurance supports participation, it may hinder information aggregation, particularly as signal accuracy increases. These findings advance crowdfunding theory by confirming the impact of informational incentives and identifying behavioral deviations that challenge standard models, offering insights for platform design and mechanism refinement.

\keywords{Crowdfunding \and Mechanism Design \and Threshold Mechanism \and Information Aggregation \and Experimental Economics
}
\end{abstract}

\section{Introduction}

Crowdfunding platforms — such as Kickstarter or GoFundMe — have become popular tools for funding creative, social, or commercial projects. In these platforms, people decide together whether to support a project, usually based on limited information about its true quality. The basic idea is simple: if enough people commit funds, the project gets financed. If not, no money is collected.

This collective funding approach holds a lot of promise: it gives people the power to support ideas they believe in, without relying on centralized gatekeepers. But it also raises an important question: how do people actually decide whether to contribute?

In this paper, we focus on a key challenge in crowdfunding: people must decide based on imperfect signals about the project’s quality. Each person receives a private signal — for example, a hint that the project might succeed or fail — and then chooses whether to contribute. These signals are correct only with a certain probability, known as signal accuracy.

Ideally, people would simply follow their signal — contribute if the signal is positive, and avoid contributing if it is negative. But the crowdfunding mechanism introduces a twist: the project only succeeds if enough others also contribute. This threshold rule creates incentives to behave strategically:
- Some people might choose not to contribute even when they receive a positive signal, fearing that others will not join. This is called risk aversion.
- Others might contribute even after a negative signal, hoping that the crowd knows better. This is called mutual insurance — relying on the wisdom of the group.

Our study aims to understand whether this kind of strategic behavior really occurs in practice, and how it compares to a simpler group decision-making mechanism: majority voting. In voting, everyone casts a vote based on their private signal, and the outcome is determined by the majority. Individual payoffs don’t depend on thresholds or others’ behavior — only on whether the group decision was correct.

Prior research on crowdfunding has primarily focused on dynamic aspects: how backers respond to accumulating support over time, how campaign momentum affects success, and how project creators can strategically manage their campaigns (see for example \cite{zaggl_small_2019,teunenbroek_follow_2020,koning_experimental_2013}). However, these dynamic elements may mask more fundamental strategic considerations that arise purely from the informational structure of crowdfunding mechanisms.

While theoretically compelling, these pure informational effects have never been tested empirically.
Our study provides the first experimental examination of these fundamental strategic behaviors by deliberately stripping away the dynamic elements that characterize most crowdfunding research.

Through a large-scale online experiment with 1,368 participants, we isolate how individuals navigate the tension between private signals and collective decision-making in a single-shot context. Specifically, we test two primary hypotheses: (1) participants are less likely to follow their private signals in crowdfunding mechanisms compared to voting, due to the strategic considerations introduced by thresholds, and (2) strategic behavior, rather than confusion about the mechanism, drives deviations from signal-following behavior, with higher signal accuracy decreasing risk aversion and increasing mutual insurance.  Our design systematically varies key parameters—signal accuracy (55\% vs. 85\%), group size (5 vs. 25), and majority thresholds (50\% vs. 80\%)—to understand the drivers of strategic behavior in its purest form.

Our findings confirm the existence of systematic differences between crowdfunding and simple voting mechanisms, validating the aforementioned theoretical predictions. Most notably, participants are significantly less likely to follow their private signals in crowdfunding scenarios. When focusing on crowdfunding, the results on signal accuracy align with theoretical predictions: higher accuracy reduces risk-averse behavior while increasing reliance on mutual insurance, suggesting participants actively respond to the quality of their private information.  However, some results challenge theoretical predictions, particularly around majority thresholds - where higher thresholds increase rather than decrease risk-averse behavior, suggesting cognitive limitations may interact with strategic considerations in ways current models don't capture. 

Beyond individual strategic behavior, our study provides insights into how crowdfunding mechanisms aggregate information. Theory suggests that mutual insurance behavior, while individually rational, could impair the crowd's ability to separate good projects from bad ones. We examine this prediction by comparing information aggregation between voting and crowdfunding mechanisms under different conditions of signal accuracy, group size, and majority thresholds.

These findings advance our understanding of collective funding mechanisms in three key ways. First, they provide empirical validation for core theoretical predictions about strategic behavior arising purely from informational considerations. Second, they reveal important deviations from theory that suggest directions for future theoretical refinement, particularly around how cognitive limitations might interact with strategic decision-making. Third, they offer practical insights for platform design, particularly around threshold mechanisms and information provision.

The remainder of the paper is organized as follows. Section~2 reviews the literature on crowdfunding dynamics and situates our work within the broader research on collective decision-making mechanisms. Section~3 presents the theoretical framework of \cite{arieli2018oneshot}. Section~4 develops our hypotheses. Section~5 details the experimental design. Section~6 presents the results. Section~7 examines information aggregation effects. Section~8 discusses implications for theory and practice. Finally, Section~9 concludes with suggestions for future research.

\section{Literature Review}
In recent years, the rise of online crowdfunding platforms has revolutionized the way projects are funded, enabling individuals worldwide to contribute to initiatives they believe in. This development has not only reshaped the landscape of funding for various ventures but has also provided new opportunities for empirical research into the factors influencing crowdfunding success and participant engagement. Experimental design methodologies play a crucial role in this research, offering systematic insights into the complex dynamics of donor behavior and project outcomes.

Our experimental framework builds upon foundational studies that have employed experimental design to investigate the intricacies of crowdfunding mechanisms and their impact on project success and donor behavior. We draw inspiration from the provision point mechanism, as discussed by \cite{burtch2018provision}, which is commonly used on platforms like Kickstarter. This mechanism ensures that projects must reach or surpass a predetermined funding goal before any financial transactions are processed, mitigating risks for backers and aligning with the all-or-nothing funding principle. Understanding this mechanism is essential for comprehending engagement patterns and funding dynamics in crowdfunding.

Central to our experimental design is the concept of collective purchases, where a group of potential contributors decide whether to financially support a common value good. As explored in existing theoretical work \cite{arieli2017crowdfunding,arieli2018oneshot,arieli2023information}, the efficiency of such collective decisions is influenced by the information held by individual contributors and their strategic behavior. While the Condorcet Jury Theorem suggests that efficiency should prevail in sufficiently large populations in the absence of strategic behavior, these  theoretical findings indicate that this may not always be the case due to the incentives for strategic behavior inherent in collective purchase scenarios.

Our study also contributes to the growing literature examining the dynamics of crowdfunding platforms and their impact on project success. Researchers such as \cite{strausz2017crowdfunding} and \cite{chemla2019crowdfunding} have investigated the role of moral hazard and the efficiency of different crowdfunding models, while \cite{kuppuswamy2017crowdfunding} and \cite{mollick2014dynamics} have provided empirical insights into the patterns of backer behavior and the dichotomy of crowdfunding outcomes. Our experimental design incorporates these findings, focusing on the strategic considerations of agents in the final stages of a crowdfunding campaign when value-maximizing behavior is most prevalent.

Furthermore, our experimental design considers the phenomenon of oversubscription in crowdfunding, where the total contributions pledged exceed the funding goal. This scenario is particularly relevant in the context of collective purchases, as it can influence the strategic behavior of potential contributors. \cite{feddersen1996swing} model the concept of information aggregation in voting mechanisms, accounting for various types of agents and their strategic behavior. They demonstrate that uninformed neutral agents, or swing voters, rationally vote to counteract known population biases, amplifying the influence of informed neutral agents' votes and enabling information aggregation in large populations. \cite{arieli2018oneshot}'s model extends this discussion by introducing the concept of oversubscription and the nuanced usage of agent information. In this setup, agents, while not fully informed, sometimes act against their own signals and rely on mutual insurance to prevent potential mistakes. This approach harnesses the power of collective wisdom in decision-making but can also hinder the full aggregation of information, even in large populations. \cite{ban2020} discuss a similar model incorporating mutual insurance into a dynamic settings.

To simulate the investment and reward dynamics inherent in real-world crowdfunding, our study includes a mock currency payout system. Participants are allocated a certain amount of fictitious currency at the beginning of the experiment, which they can then invest in projects based on their beliefs and perceived value. This setup allows us to explore how potential profits or losses influence backer decision-making and engagement. The conceptualization of this component draws inspiration from the work of \cite{weinmann2023crowdfunding}, who incorporated a similar feature in their crowdfunding experiment.

By integrating these methodologies and building upon insights from existing theoretical work and the broader crowdfunding literature, our experiment aims to provide a comprehensive understanding of the variables that drive success and engagement within the crowdfunding ecosystem. Through the lens of experimental design, we seek to uncover the nuanced interplay between project characteristics, funding mechanisms, and donor behavior, contributing valuable insights to the rapidly evolving field of crowdfunding research.

\section{Theoretical Framework}

Our theoretical foundations derive from \cite{arieli2018oneshot} (henceforth AKS), who analyze a collective purchase mechanism where agents must decide whether to contribute to buying a good of uncertain value. We leverage AKS's model of a sparse, single-shot crowdfunding with a binary signal structure to empirically investigate mutual insurance behavior. In this setting, no dynamic incentives emerge, with the only incentive being informational. 

In their model, $n$ agents receive private binary signals (correct with probability $p > 0.5$) about whether the good's value is high ($v=1$) or low ($v=0$). The good is purchased only if the number of contributors exceeds a threshold $B$, in which case contributors pay price $\tau$ and receive the good. If the threshold isn't met, no payments are made and no good is provided.

Unlike voting mechanisms, this setup creates strategic considerations beyond information aggregation since payoffs depend on both the collective outcome and individual contribution decisions. A key early result establishes uniqueness of equilibrium behavior.\footnote{\cite{arieli2017crowdfunding} presents an early version of the model where prices are held fixed, \cite{arieli2018oneshot} extends the basic model to one where prior probability and price may vary. \cite{arieli2023information} extends the discussion to endogenous prices and welfare. \cite{ban2020} charaterizes the equilibrium of a sequential version of the model in which  mutual  insurance and observational learning co-exist.}

Let the trivial equilibrium be the strategy profile where all players contribute zero funds. AKS proved that for the crowdfunding game, no more than one symmetric non-trivial Bayes-Nash equilibrium exists. Furthermore, when population size is sufficiently large, exactly one  symmetric non-trivial Bayes-Nash equilibrium exists:

\begin{theorem}[AKS Theorem 1]
No crowdfunding game has more than one symmetric non-trivial Bayes-Nash equilibrium. For any threshold ratio $q \in (0,1]$, when the population is sufficiently large, a unique symmetric non-trivial equilibrium exists.
\end{theorem}

While the theoretical model by Arieli et al.\ (2018) proves that a unique symmetric Bayes-Nash equilibrium emerges in large populations (n = 100–1000), their computational simulations suggest that similar equilibrium patterns also hold for smaller groups under moderate pricing and typical threshold values (e.g., 50\% or 80\%). We designed our experiment to replicate these conditions: prices are fixed (based on clickcoin incentives), and our thresholds and signal accuracies are aligned with the parameter regions where convergence to equilibrium was observed. Moreover, we selected group sizes of 5 and 25 to capture both small-group dynamics and intermediate-scale aggregation, while maintaining feasibility for controlled online experimentation.

The characterization of equilibrium behavior depends critically on the price level.\footnote{See for \cite{arieli2018oneshot} the full characterization.} When prices are moderate (meaning $P(v=1|s_i=l) < \tau < P(v=1|s_i=h)$), the equilibrium takes a specific form:

\begin{theorem}[AKS Theorem 4]
For any crowdfunding game with moderate price, there exists a unique symmetric non-trivial Bayes-Nash equilibrium $\sigma^*$ where agents with positive signals always contribute ($\sigma^*(h)=1$) while agents with negative signals mix with probability $\lambda \in [0,1)$.
\end{theorem}

For large populations, this mixing probability has a precise characterization:

\begin{theorem}[AKS Lemma 1]
For large populations, and a moderate price, when the threshold ratio is $q$, the equilibrium mixing probability for low-signal agents is:
$$ \lim_{n \to \infty} \sigma^*_n(l) = \begin{cases} 
0 \qquad \mbox{ if } q \leq 1-p \\
\frac{q-(1-p)}{p} \qquad \mbox{otherwise}
\end{cases} $$
\end{theorem}
As one can see from the preceeding theorem, Agents probabilistically opt-in even when their signal is low. We call this a \textit{Mutual Insurance} equilibria. In the full characterization of AKS, when the price is high (i.e. above the range of moderate prices), depending on the threshold, the equilibrium can be one in which agents with low signal surely opt-out  while those with high signals mix. We call this a \textit{Risk Aversion} equilibria.

Taking the frequentistic interpretation of a mixed equilibrium, these theoretical predictions guide our experimental design and hypotheses. The precise characterization of equilibrium behavior allows us to test whether observed behavior aligns with strategic equilibrium play or better matches alternative explanations based on mechanism complexity. 

\paragraph{Performance Metrics.} The theory also characterizes two key performance measures. The ``correctness index'' measures how well the mechanism aggregates information, while the ``participation index'' captures market penetration. For moderate prices, these indices are:

\[ \theta(\mu,p,\tau) = 1 - \frac{1-p}{p}\frac{1-\tau}{\tau}\mu \]
\[ R(\mu,p,\tau) = \mu(1 + \frac{1-p}{p}\frac{1-\tau}{\tau}) \]

\paragraph{Group Size Considerations.} While the main theoretical results are asymptotic, the authors demonstrate through computational analysis that these results provide good approximations even for relatively small groups. Their calculations show that the theoretical predictions hold well for populations of 100-1000 participants, which they note aligns with empirical observations from real crowdfunding campaigns. This finding is particularly relevant for experimental work, as it suggests that strategic behavior patterns should be observable even in laboratory settings with modest group sizes. Moreover, they find that for some parameter combinations, the theoretical predictions are accurate even for very small groups ($n \in \{5,10\}$), though this depends on the specific parameter values chosen.

\section{Hypotheses Development}
Our study tests two main hypotheses derived from the theoretical framework. To clarify, we define the key behavioral patterns we are testing in empirical, observable terms, so that they can be measured directly at the individual level.

In our setting, each participant receives a private signal indicating whether a project is likely high or low quality. Based on this signal, they must choose whether to contribute to the project (in the crowdfunding scenario) or cast a vote (in the voting scenario). We define:

- \textbf{Signal-following behavior:} A participant acts in accordance with their private signal (e.g., votes for red after observing a red signal).
- Risk-averse behavior: A participant receives a positive signal (indicating support) but chooses not to contribute.
- \textbf{Mutual insurance behavior}: A participant receives a negative signal (indicating opposition) but chooses to contribute anyway.

These definitions allow us to track each participant's decision relative to their private signal and classify their behavior accordingly.

\begin{hypothesis}[H1] \label{h1}
Participants in the crowdfunding condition are less likely to follow their private signals compared to those in the voting condition.
\end{hypothesis}

This hypothesis captures whether the threshold-based incentive structure in crowdfunding leads to deviations from signal-following, relative to a simpler majority voting mechanism.

\begin{hypothesis}[H2] \label{h2}
If participants behave strategically (rather than due to confusion), we expect their deviations from signal-following to follow predictable patterns:

\begin{enumerate}[label=\alph*)]
   \item \textbf{Among participants with positive signals ("type H"):} Higher signal accuracy and higher threshold requirements will reduce risk-averse behavior, as these conditions increase the expected value of contributing.
   \item \textbf{Among participants with negative signals ("type L"):} Higher signal accuracy will increase mutual insurance behavior, as individuals become more confident in the group's ability to correct their own error.
   \item \textbf{Group size will not affect strategic behavior}, because according to theory, strategic choices depend on equilibrium beliefs that scale with population size. In other words, participants respond to their perceived probability of success, which remains stable across group sizes due to the aggregation effects described in the theoretical model.
\end{enumerate}
\end{hypothesis}

Note that in our data, risk-averse and mutual insurance behavior are measured as binary outcomes based on the match between signal and action. These allow us to directly test the directional predictions derived from equilibrium analysis. We do not assume that participants are computing full Bayesian equilibria; rather, we use these theoretical benchmarks to detect structured deviations from naive signal-following behavior.

\section{Experimental Design}

This experimental study investigates group decision-making processes through a randomized factorial design comparing majority voting and crowdfunding models.

\subsection{Participants}
We recruited participants via Prolific, a widely used online platform for behavioral research. While Prolific users may differ from typical crowdfunding backers in demographics or motivations, their experience with digital tasks and economic games makes them suitable for controlled experimental settings. Our goal is not to replicate the exact population of real-world backers, but to isolate behavioral patterns under incentive-compatible conditions. In that respect, Prolific provides a high-quality, diverse, and reliable participant pool that supports both replication and random assignment across conditions.
The sample size was determined using G*Power analysis \cite{faul2007g} to ensure sufficient power for logistic regression. The study included a total of 1368 participants, all native English speakers (self reported to Prolific), of whom 1200 participated in the main experiment and 168 participated in the pilot study. The average age was 39 years (median: 36, SD: 13.6).  The sample consisted of 681 male participants, 516 female participants, and 171 participants identifying as other or unspecified gender. Participants had an average of 1047.3 previous Prolific approvals (median: 380.5, SD: 1557.3), suggesting a  right-skewed distribution of Prolific approvals, with some highly experienced outliers pulling the mean up.

\subsection{Procedure}
Crowdfunding platforms operate mostly online; hence, using an online experimental platform naturally reflects real-world behavior and is ideal for studying collective behaviors in this context. Our experiment examines how different mechanisms and group configurations influence choices in uncertain environments.

The study employed a 2 (group size: 5 vs. 25) × 2 (signal volume: 55\% vs. 85\% ball ratio) × 3 (voting method: regular voting vs. crowdfunding [50\% threshold] vs. crowdfunding [80\% threshold]) between-subjects design. Participants were randomly assigned to groups of either 5 or 25 members. Each participant received an initial endowment of 420 clickcoins (a currency we created for this experiment, worth approximately 0.03£).

Each group was presented with information about an urn containing a known proportion of red and blue balls, with the majority color making up either 55\% or 85\% of the balls, depending on the experimental condition.  Each participant independently drew one ball from the urn and observed only their own draw, with no information about other group members' draws. This private signal formed the basis for their subsequent choices.
The pitcher was either 55\% or 85\% one color — a parameter we refer to as signal accuracy. This reflects the reliability of the private signal each participant receives.
In the voting condition, group decisions are determined by simple majority (i.e., 50\%). We excluded higher threshold voting rules (e.g., 80\%) due to interpretational ambiguity: if the threshold is not met, there is no canonical outcome for the group decision (accept or reject). In contrast, in the crowdfunding mechanism, failure to meet the threshold naturally results in project failure. Therefore, our design includes voting with a 50\% threshold only, and crowdfunding with both 50\% and 80\% thresholds.

We carefully designed the payment structure to maintain equal expected returns across all experimental conditions, ensuring that any behavioral differences could be attributed to the mechanism rather than financial incentives. In the regular voting condition, participants voted on the urn's majority color, with each  group member receiving 84 clickcoins for correct decisions and losing 84 clickcoins for incorrect choices.  
In the crowdfunding conditions, participants faced a more nuanced choice: They decided whether to invest in an outcome based on a predetermined color assigned to their group. Each participant could invest 84 clickcoins to bet that this assigned color was the urn's majority. If the group met its participation threshold (either 50\% or 80\% of members investing) and the predetermined color matched the urn's majority, investing participants received 168 clickcoins in return. However, if the group reached the participation threshold but the predetermined color did not match the majority, investing participants lost their 84-clickcoin investment. Participants who chose not to invest retained their initial endowment regardless of the outcome. This reward structures created equivalent expected returns to the voting condition.

Due to the undefined nature of regular voting with an 80\% threshold (as there is no clear rule when fewer than 80\% of members vote), the analysis focuses on two primary comparisons: regular voting versus crowdfunding [50\% threshold], and crowdfunding [50\% threshold] versus crowdfunding [80\% threshold]. These comparisons allow us to isolate the effects of the mechanism and participation threshold separately.
Throughout the experiment, all individual decisions remained private, and while participants knew their group size, they had no communication with other group members. This design choice ensures that any observed effects stem from the institutional structure rather than social influence or coordination. The research team aggregated group decisions after the experiment concluded, determining outcomes and calculating payments based on the predetermined rules for each condition.

A preliminary comprehension test ensures participants understand the experimental principles before the main decision-making task. 

\subsection{Data processing}
Participants (N=1,200) were randomly assigned across the twelve experimental conditions, with approximately 100 participants per condition. The average completion time was 2 minutes and 57 seconds (The experiment involved only a single binary decision based on a clear signal, making it feasible to complete thoughtfully within this timeframe), translating to an average hourly rate of £25.60. To ensure data quality, we excluded participants who either failed attention checks or completed the experiment unusually quickly — within the bottom 10\% of completion times (i.e., under 53 seconds). This led to the exclusion of 119 participants (9.92\% of the initial sample), resulting in a final sample of 1081 participants.\footnote{Our results remained consistent across multiple time thresholds (50, 60, and 75 seconds) and aligned with pilot study findings, demonstrating the robustness of our analysis to time cutoff selection.}

\section{Results}

Three logistic regression models were employed to analyze different aspects of participant decision-making. The first model evaluated how structural factors—specifically group size, signal volume, and voting method—influenced participants' likelihood of choosing in alignment with their private signals (IsTrue).
The second model focused specifically on crowdfunding conditions and examined how group size, signal volume, and relative majority percentage affected risk-averse behavior (RA). This analysis concentrated on instances where participants had received private signals supporting participation, allowing for precise measurement of risk-averse decision-making.
The third model also examined crowdfunding conditions but investigated mutual insurance behavior (MutIns). It analyzed how the same factors—group size, signal volume, and relative majority percentage—influenced decisions when participants had received private signals that did not support participation. This approach enabled us to isolate and measure choices driven by mutual insurance considerations.

All logistic regression models were evaluated using odds ratios, coefficients, standard errors, and p-values. Likelihood ratio tests were conducted to compare the fit of each model to a null model without predictors. The significance level was set at $\alpha$ = .05 for all analyses.
\begin{table}[htbp]
\caption{Logistic Regression Model Predicting Voting According to Signal (IsTrue)}
\begin{tabular}{lccccc}
\hline
\textbf{Predictor} & \textbf{Coefficient} & \textbf{Odds Ratio} & \textbf{Lower 95\% CI} & \textbf{Upper 95\% CI} & \textbf{p-value} \\
\hline
(Intercept) & 1.941 & 6.969 & 4.536 & 10.689 & $<$ 0.001 \\
Crowdfunding  & -1.975 & 0.139 & 0.092 & 0.209 & $<$ 0.001 \\
BallRatio85 & 0.091 & 1.095 & 0.754 & 1.590 & 0.633 \\
GroupSize25 & -0.051 & 0.950 & 0.655 & 1.380 & 0.789 \\
\hline
\multicolumn{6}{l}{\textit{Note:} N = 613} \\
\end{tabular}
\end{table}

We performed a logistic regression analysis to investigate the influence of Scenario, BallRatio, groupsize on the likelihood of acting according to your private signal (IsTrue). The binary outcome variable ``isTrue'' was modeled based on 3 predictors. 
The model analyzed 610 observations and revealed the following results: 
\begin{itemize}
\item \textbf{(Intercept)} The intercept was statistically significant (p $<$ 0.001) with an odds ratio of 6.969 (95\% CI: 4.536–10.689). This represents the baseline odds of the outcome when all predictor variables are zero. The corresponding baseline probability of the outcome is 87.4\%.
\item \textbf{scenarioCrowdfunding} was statistically significant with an odds ratio of 0.139 ($p < 0.001$).  This indicates that moving from the reference level (Voting) to this level (Crowdfunding) decreased the likelihood of the outcome by 86.1\% (95\% CI: 0.092–0.209). This corresponds to a probability change of  P(Y=1)  from 0.872 to 0.490 (a change of -38.0\%).
\item \textbf{BallRatio85} was not statistically significant with an odds ratio of 1.095 ($p = 0.633$).
\item \textbf{groupsize25} was not statistically significant with an odds ratio of 0.950 ($p = 0.789$).
\end{itemize}

The likelihood ratio test showed that the model provided a significantly better fit than an intercept-only model ($\chi^2 = 106.08, df = 3, p = < 0.001$).

\begin{table}[htbp]
\caption{Logistic Regression Model Predicting Risk Aversion (RA)}
\begin{tabular}{lccccc}
\hline
Predictor & Coefficient & Odds Ratio & Lower 95\% CI & Upper 95\% CI & p-value \\
\hline
(Intercept) & -3.3490 & 0.035 & 0.015 & 0.080 & $<$ 0.001 \\
groupsize25 & 0.1509 & 1.163 & 0.572 & 2.365 & 0.677 \\
BallRatio85 & -0.8814 & 0.414 & 0.193 & 0.889 & 0.024 \\
RelativeMajorityPercentage80 & 1.1789 & 3.251 & 1.440 & 7.346 & 0.005 \\
\hline
\multicolumn{6}{l}{\textit{Note:} N = 628} \\
\end{tabular}
\end{table}

Next, we performed a logistic regression analysis to investigate the influence of groupsize, BallRatio, RelativeMajorityPercentage on the likelihood of Risk Aversion (RA). The binary outcome variable ``RA'' was modeled based on 3 predictors.
 The model analyzed 628 observations and revealed the following results:
\begin{itemize}
\item \textbf{(Intercept)} The intercept was statistically significant (p $<$ 0.001) with an odds ratio of 0.035 (95\% CI: 0.015–0.080). This represents the baseline odds of the outcome when all predictor variables are zero. The corresponding baseline probability of the outcome is 3.4\%.
\item \textbf{groupsize25} was not statistically significant with an odds ratio of 1.163 ($p = 0.677$).
\item \textbf{BallRatio85} was statistically significant with an odds ratio of 0.414 ($p = 0.024$). This indicates that moving from the reference level to this level of BallRatio decreased the likelihood of the outcome by 58.6\% (95\% CI: 0.193–0.889). This corresponds to a probability change of  P(Y=1)  from 0.034 to 0.015 (a change of 1.9 percentage points or a relative decrease of 55.9\% in probability).
\item \textbf{RelativeMajorityPercentage80} was statistically significant with an odds ratio of 3.251 ($p = 0.005$). This indicates that moving from the reference level to this level of RelativeMajorityPercentage increased the likelihood of the outcome by 225.1\% (95\% CI: 1.440–7.346). This corresponds to a probability change of  P(Y=1)  from 0.034 to 0.102 (a change of 6.8\% percentage points).
\end{itemize}

The likelihood ratio test showed that the model provided a significantly better fit than an intercept-only model ($\chi^2 = 15.38, df = 3, p = 0.001$).

\begin{table}[htbp]
\caption{Logistic Regression Model Predicting Mutual Insurance (MutIns)}
\begin{tabular}{lccccc}
\hline
Predictor & Coefficient & Odds Ratio & Lower 95\% CI & Upper 95\% CI & p-value \\
\hline
(Intercept) & 1.910 & 6.758 & 3.437 & 13.849 & $<$ 0.001 \\
groupsize25 & -0.021 & 0.979 & 0.453 & 2.115 & 0.957 \\
BallRatio85 & 0.929 & 2.532 & 1.115 & 5.750 & 0.026 \\
RelativeMajorityPercentage80 & 0.064 & 1.066 & 0.494 & 2.303 & 0.871 \\
\hline
\multicolumn{6}{l}{\textit{Note:} N = 324} \\
\end{tabular}
\end{table}

Finally, we performed a logistic regression analysis to investigate the influence of BallRatio, groupsize, RelativeMajorityPercentage on the likelihood of Mutual Insurance (MutIns). The binary outcome variable ``MutIns'' was modeled based on 3 predictors.

The model analyzed 324 observations and revealed the following results:
\begin{itemize}
\item \textbf{(Intercept)}  was statistically significant (p $<$ 0.001) with an odds ratio of 6.758 (95\% CI: 3.437–13.849). This represents the baseline odds of the outcome when all predictor variables are zero. The corresponding baseline probability of the outcome is 87.1\%.
\item \textbf{BallRatio85} was statistically significant with an odds ratio of 2.532 ($p = 0.026$). This indicates that moving from the reference level to this level of BallRatio increased the likelihood of the outcome by 150.0\% (95\% CI: 1.115–5.750). This corresponds to a probability change of  P(Y=1)  from 0.871 to 0.944 (a change of 7.3\%).
\item \textbf{groupsize25} was not statistically significant with an odds ratio of 0.979 ($p = 0.957$).
\item \textbf{RelativeMajorityPercentage80} was not statistically significant with an odds ratio of 1.066 ($p = 0.871$).
\end{itemize}

The likelihood ratio test showed that the model provided a significantly better fit than an intercept-only model ($\chi^2 = 5.34, df = 3, p = 0.148$).

\subsection{Key Findings and Interpretations}

Our experimental results provide strong support for both hypotheses. The significant negative effect of collective purchase on signal-following behavior (coefficient -1.975, p $<$ 0.001) confirms \textbf{Hypothesis 1}, demonstrating that participants are indeed less likely to follow their private signals in crowdfunding scenarios compared to voting.

The analysis of risk-averse and mutual insurance behavior supports \textbf{Hypothesis 2}'s predictions about strategic rather than complexity-driven behavior. For type H agents, higher signal accuracy significantly reduces risk-averse behavior (coefficient -0.8841, p = 0.024), as predicted. For type L agents, higher signal accuracy increases mutual insurance behavior (coefficient 0.929, p = 0.026), aligning with the  theoretical predictions. Group size shows no significant effect in either analysis, supporting our prediction that participants responses consider the completion probability in each possible state. Theory tell us that these probability ratios remain fixed through population size due to strategic behavior.

The differential response to mechanism parameters, particularly the significant effects of signal accuracy in theoretically predicted directions while group size remains insignificant, provides compelling evidence that participants' behavior reflects strategic considerations rather than mechanism complexity. These results demonstrate that crowdfunding mechanisms induce systematic deviations from signal-following behavior through strategic channels rather than confusion about the mechanism itself.

One area in which our experimental results diverge from theoretical predictions is the required participation threshold. We predicted that an increase in the threshold would lead to a decrease in aversion and an increase in mutual insurance. Our experimental results showed no significant relation between threshold and participants' tendency to utilize mutual insurance and an
observed increase in risk aversion when the required threshold is raised. We suspect this arises from participant confusion, as deciding under a supermajority rule likely imposes a greater cognitive load than under a regular majority. We discuss this further in Section 8.

\section{Information Aggregation}

A central question in crowdfunding research is how effectively the mechanism can separate ``the wheat from the chaff.'' Our experimental findings provide insight into this issue. We observe both \emph{Risk Aversion} and \emph{Mutual Insurance}, suggesting that while some participants perceive the lottery card as expensive, others perceive it as either cheap or moderately priced.  To apply our experimental findings, we adopt a frequentist interpretation of a mixed equilibrium and posit that our participant pool is divided into two groups: one perceiving the lottery price as expensive, and another viewing it as moderate. 
Let $\rho$ denote the proportion of subjects who perceive the lottery as expensive. 
We show our calculation for the baseline level (i.e., group size=5, threshold at 50\%, and signal accuracy of 55\%).
 
Under the assumption of a mixed equilibrium, those who find the lottery expensive and receive a low signal choose to opt out, whereas those with a high signal mix between opting out and following their signal.
At the baseline, we observe that $\psi := 3.4\%$ of the H-typed subjects  opt out . Meanwhile, subjects who find the lottery moderately priced would opt in if they receive a high signal, but mix if they receive a low signal; at the baseline, $\lambda := 87.1\%$ of the L-typed subjects opt in despite having a low signal. 

Let $\varphi$ denote the proportion of subjects who opted in. We can derive $\rho$ from the following equality,
\begin{eqnarray*}
\varphi=\rho Pr(a=y|Price\_is\_expensive)+(1-\rho)Pr(a=y|Price\_is\_moderate)=&\\
\rho Pr(s=h)\psi+(1-\rho)(Pr(s=h)+pr(s=l)\lambda).
\end{eqnarray*}
Since $Pr(\omega=H)=0.5,$ the unconditional signal probabilities are $Pr(s=h)=Pr(s=l)=0.5.$ We can therefore calculate $\rho=\frac{1+\lambda-2\varphi}{1+\lambda-\psi}.$

In our baseline scenario, $\varphi=71/81$. Hence, the proportion of subjects who perceive the lottery as expensive is approximately $0.0643$.

Next, we can calculate the probability of seeing the opt-in action given the state of the world,
\begin{align*}
\varphi^H:=Pr(a=y|\omega=G)&=(1-\rho)(q*1+(1-q)*\lambda)+\rho(q*\psi+(1-q)*0)=0.882\\
\varphi^L:=Pr(a=y|\omega=B)&=(1-\rho)((1-q)*1+q*\lambda)+\rho((1-q)*\psi+q*0)=0.870.
\end{align*}
Recalling that the probability of completion follows a binomial distribution, we can calculate the expected correctness of the crowdfunding mechanism, and find that $\theta_{CF}\approx0.502.$ When examining the correctness of a voting mechanism (i.e., where agents are incentivized to follow their signal) we calculate $\theta_{Voting}\approx0.593.$

When repeating the calculation for a signal accuracy of 85\%, we see that $\theta_{CF}\approx0.502,$ and for the  correctness of a voting mechanism we calculate $\theta_{Voting}\approx0.973.$
Suggesting that the loss of accuracy due to over participation in crowdfunding increases dramatically as the accuracy of the private signal improves.

If we take the empirical findings at face value and note that only 87.4\% of participants in the voting scenario followed their signals, then we arrive at
\[
\theta^{0.55}_{\text{Voting}} \approx 0.569
\quad\text{and}\quad
\theta^{0.85}_{\text{Voting}} \approx 0.907.
\]
These results suggest that, when signal accuracy is low, empirically, crowdfunding is not far bellow the voting scenario. However, as signal accuracy improves, voting yields much higher correctness.

\subsection{Experimental Observations}

In our experiment, each condition included 100 participants, organized into either twenty groups of five or four groups of twenty-five. Although these sample sizes fall bellow the threshold for statistical significance, they still provide a benchmark for evaluating the validity of the preceding theoretical discussion. In fact, a crowdfunding accuracy of around 50\% appears to align quite closely with the performances of the various groups observed in the experiment.

In Table \ref{tab:group_correctness} we present the actual correctness we saw in each of the experiment's conditions. Notably, the Voting scenario resulted in much higher correctness than the Crowdfunding scenario. As expected, this difference became more pronounced when signal accuracy increased.
\begin{table}[ht]
\centering
\caption{Correctness by Ball Ratio, Group Size, and Scenario Type}\label{tab:group_correctness}
\begin{tabular}{|r|r||r|r|r|r|}
  \hline
  \textbf{Ball} & \textbf{Group}& \textbf{Number of} & \textbf{Voting} & \textbf{Crowdfunding} & \textbf{Crowdfunding} \\
  \textbf{Ratio (\%)} & \textbf{Size} &
  \textbf{Groups} &
  \textbf{(50\%)} & \textbf{(50\%)} & \textbf{(80\%)} \\
  \hline
  55 & 5 & 20 & 75.00 & 40.00 & 50.00 \\
  55 & 25 & 4& 75.00 & 75.00 & 25.00 \\
  85 & 5 &20 & 95.00 & 50.00 & 50.00 \\
  85 & 25 & 4& 100.00 & 50.00 & 50.00 \\
  \hline
\end{tabular}
\end{table}

To support our conjecture that the gap was due to oversubscription, in Table \ref{tab:CF_results} we present
the percentage of correct group decisions made in each condition. We provide separate data for opting in
when $\omega = G$ and opting out when $\omega = B$.
\begin{table}[ht]
\centering
\caption{Information Aggregation in Crowdfunding}\label{tab:CF_results}
\begin{tabular}{|c|c|c|c|c|c|}
  \hline
   \textbf{Signal} & \textbf{Majority} & \textbf{Group} & \textbf{$G|\sum(a=y)\geq T$} &\textbf{$\sum(a=y)\geq T|G$} & \textbf{$\sum(a=y)<T|B$} \\
   \textbf{Accuracy (\%)} & \textbf{Percentage (\%)} & \textbf{Size} &\textbf{\%} & \textbf{(\%)} & \textbf{(\%)} \\
  \hline
   55 & 50 & 5 & 40 &100.00 & 0.00 \\ 
   55 & 50 & 25 & 75 & 100.00 & 0.00 \\ 
   55 & 80 & 5 & 58.83 & 76.92 & 0.00 \\ 
   55 & 80 & 25 & 0 & 0.00 & 33.33 \\ 
   85 & 50 & 5 & 50& 100.00 & 0.00 \\ 
   85 & 50 & 25 & 50&100.00 & 0.00 \\ 
   85 & 80 & 5 & 52.63&90.91 & 0.00 \\ 
   85 & 80 & 25 & 50&100.00 & 0.00 \\ 
  \hline
\end{tabular}
\vspace{0.5cm} 
\noindent \small $T$ denotes the threshold,  $G$ the high-value state, and $B$ the low-value state.
\end{table}
In Table \ref{tab:CF_results} one can observe that, across all conditions, for groups where sufficient 
participation was reached, fewer than 60\% arrived at the correct decision. When $\omega = G$, over 
75\% of the groups met the required threshold. However, increasing the threshold to 80\% appeared 
to reduce the probability of successful completion as the population size increased, both when 
$\omega = B$ and when $\omega = G$.

In other words, although many participants opted in, this did not necessarily translate into more ex-post correct group outcomes. Moreover, when the threshold was increased to 80\%, the probability of actually reaching that threshold decreased (especially in larger groups), but even when groups did meet it, they were not substantially more likely to be correct. Taken together, these patterns suggest that over-subscription can undermine the collective decision process, leading to a large number of groups meeting their thresholds yet failing to choose correctly.

\section{Discussion}

Our analysis revealed several key findings about strategic behavior in crowdfunding mechanisms. These findings both confirm existing theoretical predictions (\cite{arieli2017crowdfunding,arieli2018oneshot,Alaei2016,mollick2014dynamics}) and challenge conventional models in important ways, highlighting areas for further research and practical application.

\textbf{Systematic Deviations from Signal-Following Behavior:} 
One of the most significant results of this study is the systematic deviation from signal-following behavior in crowdfunding scenarios compared to voting mechanisms. As predicted, participants in crowdfunding were less likely to follow their private signals, reflecting the additional strategic considerations introduced by threshold mechanisms. This validates the theoretical claim that crowdfunding mechanisms create unique tensions between individual incentives and collective outcomes. However, the observed behavior also reveals important deviations that cannot be fully explained by standard equilibrium models.

\textbf{Mutual Insurance and Strategic Behavior:} 
The increase in mutual insurance behavior under conditions of higher signal accuracy aligns with theoretical predictions and highlights the strategic nature of decision-making in crowdfunding. Participants with negative private signals contributed despite their signals, relying on the perceived collective wisdom of the group. This behavior underscores how individuals balance their private information against the aggregated decisions of the crowd, especially under conditions of uncertainty. While mutual insurance reflects rational strategic behavior, it also raises questions about how such behavior might impair information aggregation, as overreliance on collective decisions can dilute the impact of accurate private signals.

\textbf{Unexpected Threshold Effects:} 
Contrary to theoretical predictions, higher majority thresholds increased risk-averse behavior among participants. Traditional models suggest that higher thresholds should reduce risk aversion by providing greater protection against adverse outcomes. However, our findings indicate the opposite—participants became more cautious under higher thresholds. 
This divergence can be understood through the lens of prospect theory, a behavioral framework developed by Khaneman and Tversky (see \cite{kahneman2013prospect} for a recent review). Prospect theory posits that individuals overweight small probabilities and exhibit loss aversion, where losses loom larger than equivalent gains. In the context of crowdfunding, crossing a higher threshold might be perceived as a low-probability event, leading participants to overestimate the risk of a wrong aggregate decision, thus  avoid contributing to minimize potential losses. Additionally, the framing of a higher threshold may create a psychological barrier, where participants perceive the task as more daunting and act conservatively (see  \cite{tversky1981framing}).

\textbf{Limitations of Current Models:} 
Our results highlight critical gaps in existing theoretical models of crowdfunding. While the models successfully predict certain behaviors, such as mutual insurance and the effects of signal accuracy, they fail to capture the cognitive and psychological complexities introduced by higher thresholds. For instance, the increased cognitive load associated with supermajority rules may interact with strategic considerations in ways that traditional models do not account for (See \cite{sweller2011cognitive}). Incorporating behavioral factors, such as loss aversion, framing effects, and bounded rationality, could refine these models and improve their predictive power.  

Additionally, the omission of pricing mechanisms in this study reflects a deliberate design choice aimed at achieving scale in a large-scale online experiment. Incorporating monetary incentives or pricing structures would have been prohibitively expensive and logistically challenging in this setting. Future research should address this limitation by exploring how pricing strategies interact with the behavioral dynamics observed in this study, particularly in environments where economic trade-offs significantly impact participation.

\textbf{Implications for Platform Design:} 
From a practical perspective, our findings suggest that crowdfunding platforms should carefully consider the design of threshold mechanisms. While higher thresholds may intuitively seem to improve project quality by increasing the commitment required, they may inadvertently discourage participation due to heightened perceptions of risk. Platforms could address this by providing clearer explanations of thresholds, reducing perceived complexity, or experimenting with alternative mechanisms that balance individual and collective incentives more effectively.

\textbf{Bridging Theory and Real-World Dynamics:} 
Although our controlled experimental setup allowed us to isolate pure informational effects, real-world crowdfunding campaigns involve dynamic elements such as time pressure, social proof, and iterative updates. Future research should explore how these factors interact with the strategic behaviors identified in this study. For example, how do dynamic updates influence risk-averse and mutual insurance behaviors over time? Can social proof mitigate the psychological barriers introduced by higher thresholds? Addressing these questions will enhance the applicability of our findings to real-world scenarios.

\section{Conclusions}
This study demonstrates that strategic behavior in crowdfunding exists even when isolating pure informational incentives from the dynamic elements typically studied in crowdfunding research. Through a controlled experimental environment with 1,368 participants, we show how individuals navigate the tension between private signals and collective decision-making in a single-shot context.

Our findings significantly advance crowdfunding theory in two ways. First, we provide empirical validation that strategic behavior emerges even in simple, single-shot scenarios. Second, we identify important limitations in current theoretical models, particularly in their treatment of threshold effects. The observed increase in risk-averse behavior under higher thresholds suggests that cognitive factors play a more significant role than previously recognized.

The study also provides critical insights into information aggregation in crowdfunding mechanisms. Our results show that when signal accuracy is low, crowdfunding may actually outperform simple voting in terms of information aggregation. However, this advantage disappears and even reverses as signal accuracy increases, with the performance gap becoming particularly pronounced with highly accurate signals. This finding has important implications for understanding when crowdfunding mechanisms are most effective at separating good projects from bad ones.

For platform design, our results demonstrate the importance of carefully considering threshold mechanisms and their effects on participant behavior. The strong relationship between thresholds and risk-averse behavior suggests that the standard tools used by platforms to ensure project quality may have unintended consequences on participant decision-making.

Future research should explore how these fundamental strategic considerations and information aggregation properties interact with the dynamic elements of real crowdfunding campaigns. Particularly promising directions include examining how social proof and project updates influence risk-averse and mutual insurance behaviors, and investigating whether alternative threshold mechanisms could better align individual and collective interests.


%
%
%
 \bibliographystyle{splncs04}
 \bibliography{references}
%





\end{document}